\documentclass[aps,prb,twocolumn,amsmath,amssymb,gallery,footinbib,superscriptaddress,floatfix]{revtex4-2}

\RequirePackage{ifthen}
\usepackage{graphicx}
\usepackage{dcolumn}
\usepackage{bm}
\usepackage{xcolor}

\usepackage[]{hyperref}
\begin{document}

\preprint{APS/123-QED}

\title{A family tree for hafnia}

\author{Nicolaie Cernov}
\email{nicolaie.cernov@list.lu}
\affiliation{Luxembourg Institute of Science and Technology (LIST), Avenue des Hauts-Fourneaux 5, L-4362 Esch/Alzette, Luxembourg}
\affiliation{Department of Physics and Materials Science, University of Luxembourg, Rue du Brill 41, L-4422 Belvaux, Luxembourg}
\author{Jorge Íñiguez-González}
\affiliation{Luxembourg Institute of Science and Technology (LIST), Avenue des Hauts-Fourneaux 5, L-4362 Esch/Alzette, Luxembourg}
\affiliation{Department of Physics and Materials Science, University of Luxembourg, Rue du Brill 41, L-4422 Belvaux, Luxembourg}
\author{Hugo Aramberri}
\email{hugo.aramberri@list.lu}
\affiliation{Luxembourg Institute of Science and Technology (LIST), Avenue des Hauts-Fourneaux 5, L-4362 Esch/Alzette, Luxembourg}

\date{\today}

\begin{abstract}
Several candidate reference phases have been proposed to discuss phase transitions and ferroelectricity in hafnia in recent years. Although these proposals comply with crystallographic requirements, a physically compelling rationale connecting parent and daughter phases is often lacking. This problem is aggravated by the absence of clearly dominant polarization switching pathways, making it difficult to formulate a robust physical criterion for identifying the relevant high-symmetry states.
Here we use first-principles calculations to show that pressure provides a simple and robust criterion for establishing physically meaningful family relationships among many hafnia polymorphs, including the monoclinic ground state and the technologically relevant ferroelectric phase, and their respective parent structures. 
Our simulations also reveal several previously unreported phases, including higher-energy structures that act as common ancestors of the widely discussed cubic and orthorhombic reference phases.
\end{abstract}
\maketitle

Hafnia is a ferroelectric oxide that challenges many established views on ferroelectrics. For one, its polarization seems to be robust against downsizing~\cite{boscke11,muller11,ali18,cheema20}, contrary to what happens in traditional ferroelectrics like perovskite oxides. Also, its longitudinal piezoelectric response is still a matter of debate and has been proposed to be of both positive and negative sign~\cite{dutta21,cheng24,lu24}. At a more fundamental level, even the sign of the polarization in its ferroelectric phase remains controversial, since it is dependent on the choice of the reference structure~\cite{choe21,wu23,qi25b}, and it has been argued that it might not be possible to determine it from the bulk structure of the material alone~\cite{mukherjee25}.

In many studies, reference structures have been proposed based primarily on crystallographic considerations, such as pseudo-symmetry relationships or group–subgroup connections between phases. However, in hafnia the atomic distortions linking different polymorphs often involve very large oxygen displacements, typically comparable to the lattice parameters themselves. Under these circumstances, purely crystallographic reasoning becomes ambiguous, since several high-symmetry configurations may serve as valid parents for a given low-symmetry structure. As a result, different candidate reference phases can lead to equally consistent symmetry-mode decompositions.

\begin{center}
  \begin{figure}
     \centering
      \includegraphics[width=\columnwidth]{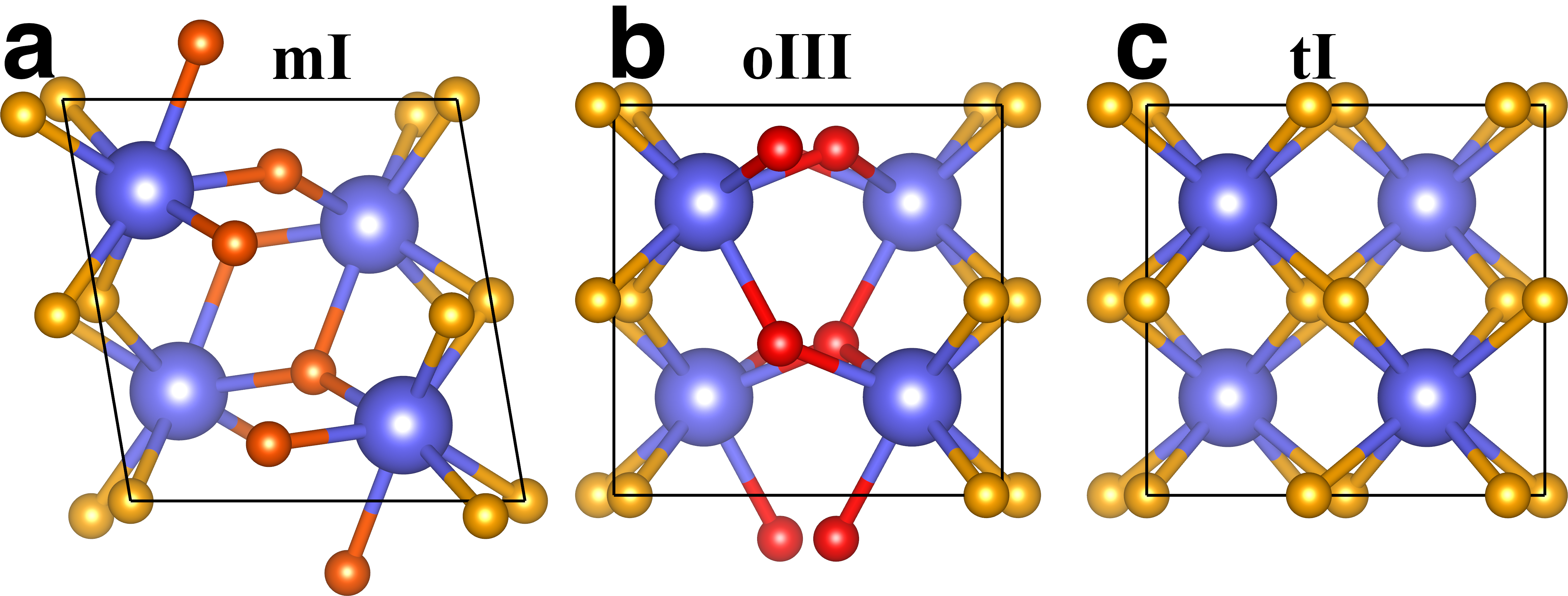}
      \caption{\textbf{Crystal structure of relevant low-energy phases of hafnia.} The inactive oxygens are coloured in yellow. In several phases, including mI (\textbf{a}) and oIII (\textbf{b}), every other oxygen column remains in 4-fold coordinated sites (inactive), while neighbouring columns undergo significant displacements and become 3-fold coordinated, resulting either in arrangements with no net dipole (in orange, \textbf{a}) or with a finite electric dipole (in red, \textbf{b}).}
      \label{fig:oxygens} 
  \end{figure}
\end{center}

This ambiguity is reflected in the variety of reference structures that have been proposed for hafnia. The tetragonal phase ("tI", with space group $P4_2/nmc$, see Fig.~\ref{fig:oxygens}\textbf{c}) has often been used~\cite{huan14,liu19,delodovici21}, although it shows no instabilities and is therefore a poor choice as a parent phase in most cases (e.g. the simplest energy expansions from a stable phase are of higher order than those from an unstable one).

Alternatively, the cubic fluorite phase ("cI"-phase, with space group $F m\bar{3}m$) has been proposed as a parent structure (initially in the intimately related compound ZrO$_2$) to describe the transition from the cI to the tI phase~\cite{mirgorodsky95}, and from the cI to the well-known orthorhombic ferroelectric phase ("oIII", with space group $Pca2_1$, see Fig.~\ref{fig:oxygens}\textbf{b}). While the cI phase is unstable, the connection to the ferroelectric phase is quite intricate - it involves first the transformation into tI followed by a second transition into an antipolar metastable phase ("oVIII", with symmetry $Pbcn$), which was claimed to then be destabilized (by epitaxial strain) and followed by a third phase transition to the oIII phase~\cite{zhou22,raeliarijaona23}. An alternative path from cI to oIII was proposed in ZrO$_2$, using an electric field to induce a first-order phase transition via tI~\cite{reyes-lilo14}.

More recently, the oVIII phase itself was proposed as a reference phase~\cite{raeliarijaona23}. However, this phase (which is a local minimum of the energy~\cite{barabash17,raeliarijaona23err}) has never been experimentally observed in hafnia,  which questions its relevance to the physics of the material.

In previous work, some of us have discussed a different orthorhombic reference~\cite{aramberri23} ("oVII", with space group $Pbcm$) from which several of the most relevant low-energy phases can be obtained by condensation of individual soft modes, including the monoclinic ground state ("mI", with space group $P2_1/c$, see Fig.~\ref{fig:oxygens}\textbf{a}), an orthorhombic phase that can be viewed as a modulated version of it ("oVI" with $Pbca$ symmetry), the ferroelectric phase, and another orthorhombic phase that is again a modulated version of the latter ("oI", also with space group $Pbca$). However, as in the case of oVIII, there is no direct experimental evidence of the physical relevance of this state, which has not yet been observed in hafnia.

To conclude this overview, let us note that yet another orthorhombic phase ("oXV" with space group $Cmme$) obtained from the oVII structure through a symmetry-restoring distortion has more recently been proposed as a reference for hafnia~\cite{qi25}. Combinations of unstable modes from this phase result in several low energy phases, including the monoclinic ground state mI and the ferroelectric phase oIII. In both cases the path involves the oVII phase as an intermediate state.

A natural possibility to define a physically meaningful reference structure for the ferroelectric phase would be the transition state involved in polarization switching. However, several distinct switching paths have been proposed for hafnia~\cite{clima14, choe21, silva23, ma23, wu23, dou24, qi25b} with critical fields of similar magnitude~\cite{hu25}. These switching paths involve large-amplitude structural rearrangements, including substantial oxygen displacements and, in some cases, anion migration. Such distortions are far from perturbative and therefore difficult to describe within Landau-like expansions around a single high-symmetry configuration. Importantly, the multiplicity of switching pathways suggests that it is impossible to describe switching in hafnia in terms of a single reference structure. In turn, this implies that we lack a criterion to establish physically meaningful parent-daughter phase connections in this compound.

In this work, we explore an alternative approach to discuss the hierarchy of polymorphs in hafnia. Rather than relying exclusively on geometrical crystallographic relationships, we analyze the structural landscape from a physical perspective based on the response of the system to hydrostatic pressure. Using first-principles calculations, we show that pressure reveals clear parent–daughter relationships among many polymorphs of hafnia and helps clarify the connections between phases in its complex energy landscape. In addition, combining phonon analysis, crystal-structure database searches, and directed exploration of the structural space, we predict several previously unreported polymorphs of this compound, including common parent structures for the cubic and oVII phases.

\section{Results And Discussion}

\begin{center}
  \begin{figure*}
     \centering
      \includegraphics[width=\textwidth]{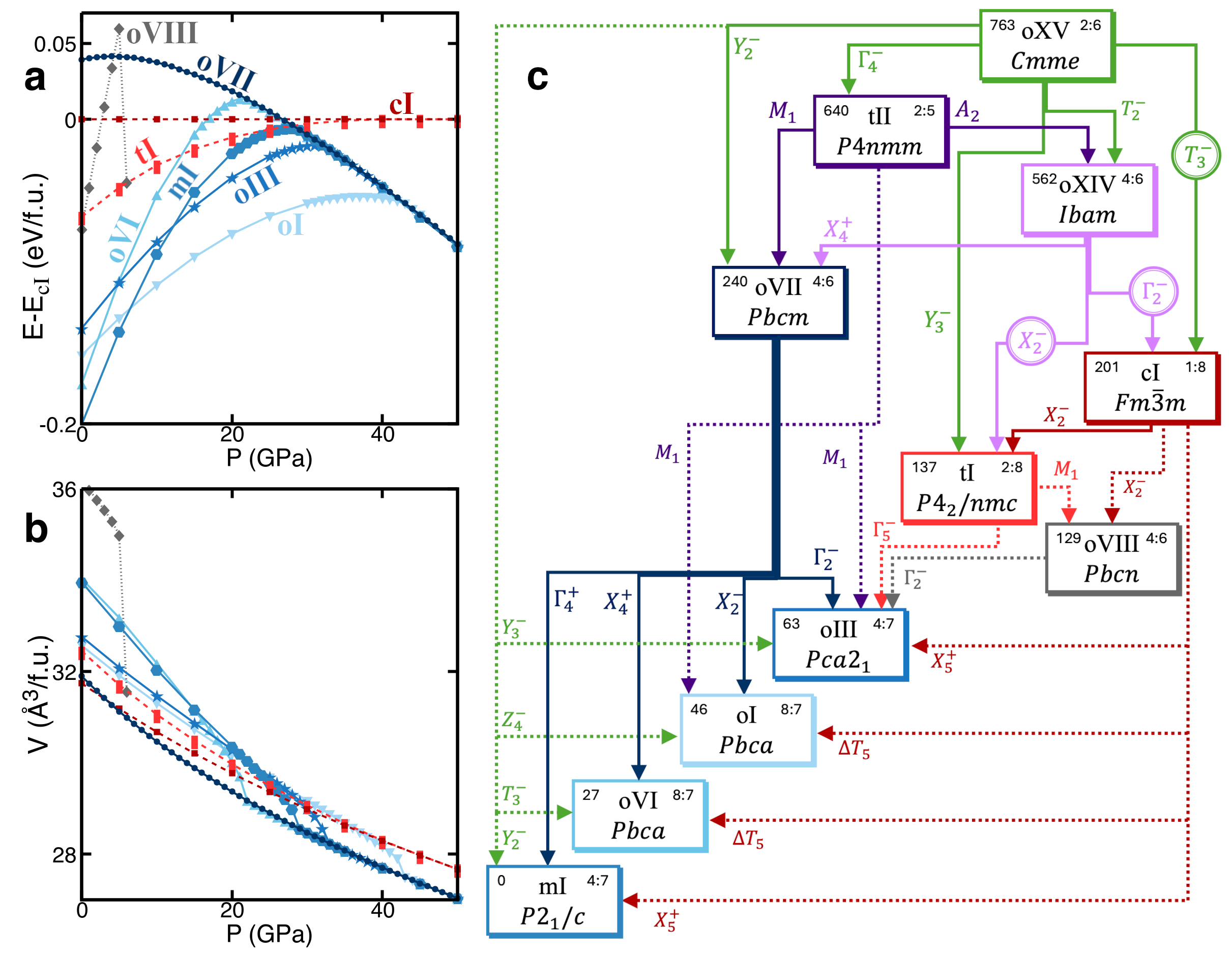}
      \caption{\textbf{Energy and volume of hafnia phases with pressure.} Evolution of energy (\textbf{a}) and volume (\textbf{b}) of the most relevant phases of hafnia as a function of pressure. Many low energy polymorphs (including the monoclinic ground state mI and the ferroelectric phase oIII) transform smoothly into the oVII phase. The tetragonal tI phase transforms smoothly into the cubic cI phase. The antipolar oVIII phase destabilizes at relatively low pressures into the tI phase. Panel \textbf{c} shows the hafnia family tree based on the transformations with pressure. Each box represents one phase of hafnia. On the upper right corner the numbers indicate the number of formula units in the primitive cell (before the colon) and the coordination number of Hf of that phase (after the colon). The number in the upper left corner shows the energy with respect to the ground state in meV/f.u. at zero pressure. The phases are ordered vertically by increasing relative energy. Phases that show instabilities (at $P=0$~GPa) are marked with solid lines with arrows pointing to the phases to which the condensation of these instabilities give rise. Dashed lines indicate parent/daughter relationships that we found via crystallographic analysis where the leading distortion is not a soft phonon mode of the parent structure. The character of the leading distortion(s) is marked on each arrow. The encircled distortions mark soft-mode condensations that result in symmetry recovery.}
      \label{fig:energyvol} 
  \end{figure*}
\end{center}

\begin{center}
  \begin{figure*}
     \centering
      \includegraphics[width=\textwidth]{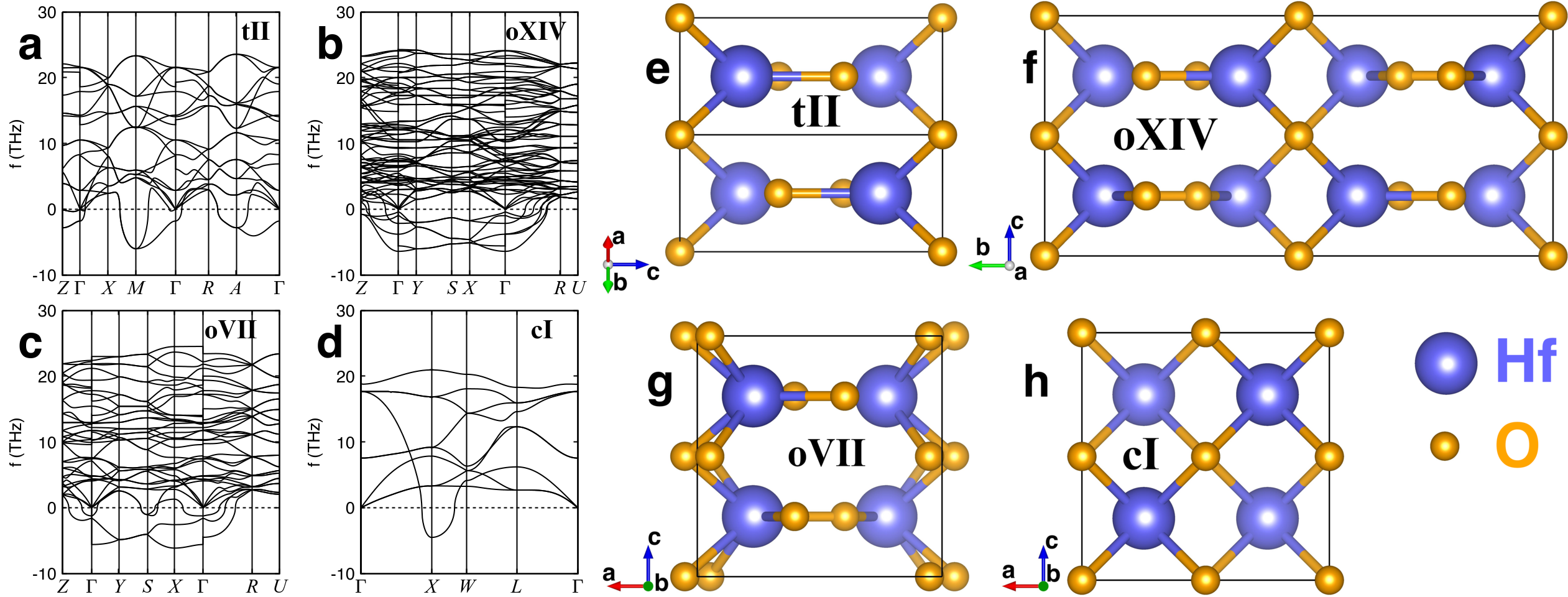}
      \caption{\textbf{Phonons and crystal structure of relevant phases of hafnia which are saddle points of the energy.} For oXIV we show the conventional unit cell; its primitive cell contains only 4 formula units.}
      \label{fig:phonons} 
  \end{figure*}
\end{center}

By following the evolution of energy and volume with pressure for the different phases of hafnia, we find that several low energy structures transform smoothly into the oVII orthorhombic phase. These include the monoclinic ground state mI and its antimodulated version oVI, the oI phase (which has been stabilized experimentally with pressure~\cite{leger93b}) and the ferroelectric phase oIII. Both the energy (see Fig.\ref{fig:energyvol}\textbf{a}) and cell volume (Fig.\ref{fig:energyvol}\textbf{b}) of these polymorphs evolve smoothly into that of oVII, indicating (i) continuous phase transitions to oVII, and (ii) that oVII is the natural parent phase for these daughters. Other phases considered show the same behaviour (see Suppl. Figs. S1\textbf{a} and S1\textbf{d}). We find that all the phases that transform smoothly into oVII can be obtained by condensing unstable phonon modes at high symmetry points of the Brillouin zone (see Fig.\ref{fig:phonons}\textbf{c})~\cite{aramberri23}. This suggests that the oVII is a particularly useful promontory of the energy landscape to understand these low energy phases.

Among the most relevant phases, only tI transforms into the cubic cI phase with pressure (two other phases of higher energy than tI also follow this behaviour, see Suppl. Fig. S1\textbf{b} and S1\textbf{e}). This phase transition also appears to be continuous. In this case also, the daughter phase tI can be obtained by condensing the instability of the parent structure, cI (see Fig.\ref{fig:phonons}\textbf{b}).

In contrast, the energy of the antipolar oVIII phase increases significantly under pressure, quickly overcoming that of oVII. Around $\sim$6~GPa oVIII becomes unstable and transforms into tI, with an abrupt change of the energy and volume, indicating a discontinuous transition upon compression (Fig.\ref{fig:energyvol}\textbf{a} and \ref{fig:energyvol}\textbf{b}).

\subsection{Common ancestors of cI and oVII}
By applying high pressures to the oVII phase we find a new tetragonal high-energy polymorph of hafnia ("tII" with space group $P4/nmm$, see Fig.~\ref{fig:phonons}\textbf{e}) where the splitting of the inactive oxygens of oVII is undone (compare Fig.~\ref{fig:phonons}e with Fig.~\ref{fig:phonons}g; see Fig~\ref{fig:oxygens} for definition of active and inactive oxygens). Phonon analysis of this tetragonal phase (see Fig.~\ref{fig:phonons}\textbf{a}) reveals that the strongest stability of tII (at the $M$ point) leads to oVII. 

The condensation of the unstable modes of tII at the $\Gamma$ and $Z$ points results in the oIV and oV phases, respectively. As already pointed out by Azevedo-Antunes and collaborators~\cite{azevedo21}, oV can be seen as a modulated version of the polar phase oIV (see also Supp. Fig. S8\textbf{g} and S8\textbf{l}). Our calculations indicate that tII can be taken as their common ancestor in this context. 

Interestingly, condensing the $A$-point instability of tII leads to yet another new phase, oXIV (see Fig.~\ref{fig:phonons}\textbf{f}). This polymorph also shows no splitting of the inactive oxygens, and the in-plane distortion of the active oxygens is modulated along the horizontal direction. The phonon spectrum of oXIV reveals several instabilities (see Fig.~\ref{fig:phonons}\textbf{b}). The strongest one at the $\Gamma$ point leads to the cI phase (Fig.~\ref{fig:phonons}\textbf{h}), while the strongest instability at the $X$ point condenses into the oVII phase (Fig.~\ref{fig:phonons}\textbf{g}). Hence, the oXIV phase can be seen as a common parent structure of both cI and oVII phases.

We also note that an instability of the oXIV phase at the $X$ point leads to tI. The other instabilities of oXIV give rise to yet more new phases of hafnia but of much higher energy, and hence of less potential interest, so we did not include them in this work. We also find that the oXIV phase smoothly transforms into the tII phase with pressure, the critical point being around 9~GPa (see Suppl. Fig. S1\textbf{c} and S1\textbf{f}). 

The recent work of Qi and Rabe~\cite{qi25} proposes a new orthorhombic reference phase with space group $Cmme$ (here "oXV") which is a saddle point of the energy. We find that this phase is of higher energy than all the phases included in this study (763~meV/f.u. above the mI ground state), and, importantly in the present context, the stability of this phase is strongly penalized with increasing pressure. Still, we find (in line with the original report of Qi and Rabe) that several key phases can be obtained by condensing individual instabilities, including the oVII and cI structures, the high-energy phases tII and oXIV and the tI phase. Thus, oXV can be seen as an ancestor of oVII and cI as much as tII and oXIV.

\subsection{Overview of structural landscape}

In Fig.\ref{fig:energyvol}\textbf{c} we summarize our findings in the form of a phase family tree for hafnia. Arrows with solid lines indicate parent-daughter relationships mediated by the condensation of one primary instability of the parent structure. Arrows with dashed lines show parent-daughter relations obtained only via crystallographic analysis, and/or where the condensation of a single phonon instability is insufficient to establish the connection. Notably, if the classification were not physically informed, the oIII phase could be seen as an offspring of either oVII, tI, oVIII, cI (as previously reported in other works) or even tII. Similarly, the tI and oVIII phases could be considered in this way as daughter phases of cI, as previously reported. However, our calculations under pressure point at the oVII as the natural vantage point of the energy landscape to describe ferroelectric hafnia. An extended version of this family tree with all the phases of hafnia here considered can be found in Suppl. Fig. S9.

In order to have a more complete overview of the structural and energy landscape of hafnia, we include in the study several other phases for which we did not find a parent/daughter connection to the phases discussed above. The evolution of their energies and volumes with pressure (together with those of the tII blood line) is shown in Suppl. Figs. S1\textbf{c} and S1\textbf{f}, where it can be seen that there are no smooth transformations to or from these other polymorphs. We note that all such phases are dynamically stable (the only exception being mV, which shows a zone-boundary instability that gives rise to the oIII* phase).

The list of all the phases studied in this work is given in Table~\ref{bigtable}, which includes their space group, energy with respect to the mI ground state, identified parent structure(s), dynamical stability, volume per formula unit, number of formula units in the primitive cell, Hf coordination and number of formula units per primitive cell. This information is also displayed in the form of a tree diagram in Suppl. Fig. S9, which is an extended version of Fig.~\ref{fig:energyvol}\textbf{c}.

Our calculations also reveal several other details that we consider noteworthy. For example, the mV phase that we found serendipitously shows a single instability that gives rise to the oIII* phase. This phase is polar and similar to the ferroelectric oIII phase, the main difference being that the oIII* presents an in-plane modulation of the active oxygens and hafnium atoms along the polar axis (see Suppl. Fig. S3). Its energy is only 44~meV/f.u. above that of the oIII, and its polarization is 0.51~C/m$^2$, smaller than the value for oIII, 0.68~C/m$^2$ (in both cases we calculated the polarization using the oVII phase as a reference). Our calculations hence indicate that in-plane distortions result in (meta)stable structures, and that these distortions can lower the total value of the polarization.  In this way, we speculate that the long wake-up often observed in real samples of ferroelectric hafnia might be partly related to undoing of in-plane distortions driven by electric field cycles.

Analogously, the oXIII phase (entry mp-776097 in the Materials Project database~\cite{materialsproject}) can be seen as an in-plane modulated version of oVIII along the direction in which the active oxygens displace (see Suppl. Fig. S4). This modulation comes at a very similar energy cost (44~meV/f.u.) than that relating oIII* and oIII.

Another interesting point is that hafnia shows at least three phases that are isostructural to phases of titania, namely rutile (tV), anatase (tIV) and brookite (oXII). These phases show a six-fold coordination for Hf, and are very strongly penalized by pressure (see Suppl. Fig. S1\textbf{c}). While the rutile phase of hafnia was already predicted by Barabash~\cite{barabash17}, the other two have not been reported to our best knowledge. Note that in Refs.~\onlinecite{raeliarijaona22} and \onlinecite{raeliarijaona23} the term brookite is used to refer to the oVI phase; however, the original brookite phase of titania (which we used as a starting point of our structural relaxation) shows a six-fold coordination of Ti (like oXII does for Hf), while the oVI phase shows a seven-fold coordination of Hf (see Suppl. Fig.~S5 for a comparison of the two crystal structures).

\begin{table*}
 \centering
 \begin{tabular}{llrcccccll}
 Phase         & Symmetry     & $E$   &S?&Parent  &$V$& f.u./cell & CN$_{\mathrm{Hf}}$ & Ref & Mineral family \& other names \\
 \hline
 mI            & $P2_1/c$     &   0.0 &Y& oVII                 &33.96&  4 & 7 &  & Baddeleyite (ZrO$_2$)\\
 oVI           & $Pbca$       &  26.6 &Y& oVII                 &34.03&  8 & 7 &  \onlinecite{ohtaka95,kersch21,azevedo22} & oI*\\
 oI            & $Pbca$       &  45.6 &Y& oVII                 &32.56&  8 & 7 & \onlinecite{jaffe05}&\\
 mIII          & $Pc$         &  49.6 &Y& oVII                 &33.16&  8 & 7 & \onlinecite{azevedo22}&\\
 oIII          & $Pca2_1$     &  63.2 &Y& oVII                 &32.76&  4 & 7 & \onlinecite{boscke11}&o-FE\\
 mII           & $P2_1/m$     &  85.3 &Y& -                    &34.06&  2 & 7 & \onlinecite{huan14,barabash17}&\\
 oXI           & $Pnma$       &  88.6 &Y& -                    &34.41&  4 & 7 & &\\
 oIII*         & $Pca2_1$     & 106.6 &Y& mV                   &32.75&  8 & 7 & &\\
 oV            & $Pnma$       & 107.7 &Y& tII                  &32.81&  4 & 7 & \onlinecite{azevedo21}&\\
 oIV           & $Pmn2_1$     & 120.4 &Y& tII                  &32.83&  2 & 7 & \onlinecite{huan14}&\\
 tIV           & $I4_1/amd$   & 121.0 &Y& -                    &41.43&  2 & 6 & &Anatase (TiO$_2$)\\
 oVIII         & $Pbcn$       & 128.6 &Y& -                    &36.18&  4 & 6 & \onlinecite{barabash17,zhou22}& Scrutinyite ($\alpha$-PbO$_2$), Columbite ((Fe.Mn)Nb$_2$O$_6$), o-AP \\
 tI            & $P4_2/nmc$   & 136.5 &Y& cI/tIII/cII/oXIV/oXV &32.44&  2 & 8 & & \\
 tIII          & $P4/nbm$     & 152.4 &N& cI                   &32.22&  4 & 8 & \onlinecite{barabash17} &\\
 cII           & $P\bar{4}3m$ & 156.4 &N& cI                   &32.16&  4 & 8 & \onlinecite{mcclellan94,wang04,barabash17} &\\
 oIX           & $Pba2$       & 158.2 &Y& oVII                 &32.27&  8 & 7 &  &\\
 oXII          & $Pbca$       & 163.8 &Y& -                    &38.59&  8 & 6 &  &Brookite (TiO$_2$)\\
 tV            & $P4_2/mnm$   & 170.5 &Y& -                    &37.13&  2 & 6 & \onlinecite{barabash17} & Rutile (TiO$_2$)\\
 oXIII         & $Pbcn$       & 173.8 &Y& -                    &36.87& 12 & 6 & &\\
 mIV           & $P2_1/c$     & 185.0 &Y& oVII                 &32.64&  8 & 7 & &\\
 rII           & $R3$         & 194.0 &Y& -                    &32.91&  4 & 7 &\onlinecite{barabash17} &\\
 cI            & $Fm\bar{3}m$ & 200.5 &N& oXIV                 &31.74&  1 & 8 & &Fluorite (CaF$_2$)\\
 oX            & $Aea2$       & 205.8 &Y& oVII                 &32.26&  8 & 7 & &\\
 oXVI          & $P2_12_12$   & 231.2 &Y& oVII                 &31.95&  4 & 7 & &\\
 mV            & $Cc$         & 234.4 &N& -                    &33.03&  4 & 7 & &\\
 oVII          & $Pbcm$       & 239.8 &N& oXIV/tII/oXV         &31.90&  4 & 6 & \onlinecite{liu14}& o-ref\\
 oII           & $Pnma$       & 275.2 &Y& -                    &29.08&  4 & 7 & \onlinecite{jaffe05} & Cottunite\\
 hI            & $P\bar{6}2m$ & 362.4 &Y& -                    &28.77&  3 & 8 & \onlinecite{yahya18} & Barringerite (Fe$_2$P)\\
 oXIV          & $Ibam$       & 561.5 &N& tII/oXV              &32.14&  4 & 6 & &\\
 tII           & $P4/nmm$      & 639.2 &N& oXV                  &30.10&  2 & 5$^\dagger$  & &\\
 oXV           & $Cmme$       & 762.7 &N& -                    &36.24&  2 & 6  &\onlinecite{qi25} &\\

 \end{tabular}
 \caption{List of hafnia phases studied. Columns 2 to 10 show the space group, the energy with respect to the ground state at $P=0$ (in meV/f.u.), whether the phase is dynamically stable (Y) or not (N), the parent phase, the volume per formula unit at $P=0$ (in \AA$^3$/f.u.), the number of formula units per primitive cell, the coordination number of Hf, bibliographic references, and other names for the phase, respectively.
 $\dagger$: the coordination number of Hf in phase tII is 5 (9) if Hf-O distances below 2.4 (2.5)~\AA\ are considered. The crystal structures are shown in Suppl. Figs. S6-S8 and the Crystallographic Information Files (CIF) are provided as supplementary data.}
 \label{bigtable}
\end{table*}

\subsection{Other considerations}
Interestingly, when we compute the dielectric permittivity for the oVII-daughter phases as a function of pressure (see Suppl. Fig. S2), we find that the $\varepsilon_{zz}$ component (where $z$ is the direction of the polar axis in the oIII phase) shows a peak for pressures just below the transition to oVII (except for the oXVI phase, where no sudden increase is observed). A peak in the dielectric response is a hallmark of proximity to a soft-mode polar instability, and is therefore expected in the oVII to oIII transformation. However, the presence of this peak close to the critical pressure in the other phases is surprising, since none of them is polar.
 Note that the soft mode of oVII that leads to oIII involves the displacement of the active oxygens in unison along the $z$ direction, which induces the polar character of this phase. The soft modes of oVII that lead to the other daughter phases similarly involve displacements of active oxygens along $z$, but without creating a net dipole. This, together with the dielectric peaks associated with their transitions to oVII, points at a possible antiferroelectric nature~\cite{rabeAFE} of the mI, oVI, oI, oIX, mIV and oX phases.
 
\section{Conclusion}

In summary, we demonstrate that pressure provides a physically transparent criterion to identify meaningful parent–daughter relationships among hafnia polymorphs, bringing significant order to the exceptionally rich energy landscape of this compound. Our analysis shows that the orthorhombic oVII phase serves as a central vantage point connecting both the monoclinic ground state and the ferroelectric oIII phase through continuous pressure-induced transformations, while the cubic cI phase is analogously directly connected to the tetragonal tI phase.

As additional discoveries, we also predict several previously unreported polymorphs, including a low-energy polar phase (oIII*) closely related to the technologically relevant ferroelectric phase. In addition, we identify higher-energy saddle-point structures (oXIV, tII and oXV) with instabilities whose condensation results in the principal parent phases oVII and cI.

Rather than seeking a unique reference phase for hafnia, our results indicate that physically meaningful structure–structure relations are best established through explicit dynamical and thermodynamic criteria. In particular, in the absence of a dominant polarization switching path, and given that proposed switching mechanisms involve large-amplitude atomic rearrangements, Landau-like expansions around a single high-symmetry configuration offer limited guidance. In this context, parent–daughter relations revealed by pressure and lattice instabilities provide a more robust and physically grounded basis for understanding hafnia's polymorphism and may help guide future theoretical and experimental investigations.

\section{Simulation methods}
We use density functional theory (DFT) calculations as implemented in the Vienna Ab Initio Simulation Package (\textsc{VASP})~\cite{kresse96,kresse99} to study the different phases of hafnia. For the exchange-correlation functional we employ the Perdew-Burke-Ernzerhof formulation for solids (PBEsol)~\cite{pbesol} of the generalized gradient approximation. The cores are treated within the projector-augmented wave (PAW) approach~\cite{blochl94}, with the following states explicitly considered: 5s, 5p, 6s, 5d for Hf; and 2s, 2p for O. 

We use a plane-wave energy cutoff of 600~eV for all calculations. Integrations in the Brillouin zone were carried out using a 6$\times$6$\times$6 Monkhorst-Pack grid~\cite{monkhorst76} for the primitive cell of the mI and for phases with similar cell sizes (oIII, tI, oVI, oVII, tIII, tVI, rII, oXVI). Similarly dense grids were employed for other phases (e.g. 3$\times$6$\times$6 for the oI, oVI, mIII, oXII and oXIV phases; 4$\times$4$\times$6 for the mIV, mV and oX phases; etc. ). 

We optimize the structures until atomic forces fall below 1~meV/\AA\ and residual stresses become smaller than 0.05~GPa. We apply pressure in the range of 0-50~GPa in steps of 5~GPa, and we reduce the step size to 1~GPa around critical pressures for the relevant phases.

We compute the phonons using the finite displacements approach as implemented in the \textsc{phonopy}~\cite{phonopy} package. To this end, we create 2$\times$2$\times$2 supercells for the oIII and similarly sized phases, which we found to be sufficiently converged. For phases with primitive cells of different sizes, we adjust the supercell size accordingly to have similarly large supercells for all phases. We consider the non-analytical contribution to the phonons in the calculations.

We use the \textsc{ISODISTORT} tool~\cite{isodistort} to obtain the symmetry-adapted distortions connecting phase pairs. 

We use \textsc{VESTA}~\cite{vesta} to prepare the figures showing crystalline structures.

\acknowledgments
We acknowledge the support of the Luxembourg National Research Fund through grants AFR/18810392/LIGHTTRAP (N.C.) and INTER/NWO/20/15079143/TRICOLOR (J.Í.-G. and H.A.). We acknowledge fruitful discussions with Binayak Mukherjee (now in Université de Liège).

\end{document}